\documentclass[sigplan, 
]{acmart}

\usepackage{booktabs} 

\newcommand{\femph}{\textit}
\newcommand{\fcode}{\texttt}

\newcommand{\sectref}[1]{\hyperref[#1]{Sec.\,\ref*{#1}}}

\definecolor{darksilver}{rgb}{0.3,0.3,0.3}

\usepackage{listings}       

\lstdefinelanguage[cpSTLC]{OCaml}[]{Caml}{
        morekeywords={concept,model,of,endc,endm,
        Definition,Fixpoint,Lemma,Prop,bool,true,false,nil,list,forall,Module,Import,End,Theorem,Proof},
}

\lstdefinestyle{normal-lst-style}{
  frame=tb,
  basicstyle=\ttfamily\scriptsize, 
  keywordstyle=\ttfamily\bfseries,
  commentstyle=\ttfamily\color{darksilver}
}

\lstnewenvironment{cpSTLC}
        {\lstset{language=[cpSTLC]{OCaml},style=normal-lst-style}}
        {}
\lstnewenvironment{cpSTLCNoFrame}
        {\lstset{language=[cpSTLC]{OCaml},style=normal-lst-style,frame=none}}
        {}

\begin{document}
\title[Generic Approach to Certified Static Checking of Modules]{Generic Approach to Certified Static Checking\\ 
of Module-like Constructs}

\author{Julia Belyakova}
\affiliation{%
  \institution{Southern Federal University}
  \city{Rostov-on-Don, Russia} 
}
\email{julbinb@gmail.com}

\copyrightyear{2017}
\acmYear{2017}
\setcopyright{acmcopyright}
\acmConference{FTFJP'17}{June 18-23, 2017}{Barcelona , Spain}\acmPrice{15.00}\acmDOI{10.1145/3103111.3104045}
\acmISBN{978-1-4503-5098-3/17/06}



\begin{abstract}
In this paper we consider the problem
of certified static checking of module-like constructs 
of programming languages. 
We argue that there are algorithms and properties related to modules
that can be defined and proven in an abstract way.
We advocate the design of a generic Coq library,
which is aimed to provide three
building blocks for each checking mechanism:
propositional, 
computable, 
and correctness proofs.
Implemented part of the library is justified by applying it
to a certified static checker of an extension of STLC.

\end{abstract}

%
%




\maketitle



%

\section{Introduction}\label{sec:intro}

Interactive theorem provers, such as Coq, Agda, or HOL, have been used for both 
\femph{mechanising formal models} of programming languages
(Featherweight Java~\cite{bib:Mackay:2012:EFJ:2318202.2318206,bib:Delaware:2011:PLT:2076021.2048113},
Scala DOT-calculus~\cite{bib:Rompf:2016:TSD:3022671.2984008},
JavaScript~\cite{bib:Bodin:2014:TMJ:2535838.2535876},
Dependent Haskell~\cite{bib:Weirich:2017:Dep-Haskell}) 
and development of certified software,
including \femph{certified compilers and interpreters}
(CompCert~\cite{bib:Blazy-Leroy-Clight-09}, 
JSRef~\cite{bib:Bodin:2014:TMJ:2535838.2535876},
CakeML~\cite{bib:Tan:2016:NVC:2951913.2951924}). 
In this context ``certified'' means that
the behaviour of an interpreter/compiled code
corresponds to the formal model
of a programming language. For example, a JavaScript interpreter JSRef
is proven to satisfy the JSCert~\cite{bib:Bodin:2014:TMJ:2535838.2535876} formalization of JavaScript.

Generally, the structure of a certified interpreter\footnote{For brevity, we only 
talk about certified interpreters from this point, but the same reasoning is
applied to certified compilers.} can be described
with three layers:
\begin{enumerate}
 \item Formal model of a programming language 
   (typing relation, operational semantics)
   defined in propositional style.
 \item Interpreter itself (static checks,
   evaluator)
   defined in terms of computable functions.
 \item Proof of correctness of the interpreter
   with regard to the formal model.
\end{enumerate}
For instance, the typechecking task 
could be lined up as follows~\cite{bib:Pierce:SF}:
\begin{enumerate}
  \item\label{enum:lab:prop} Typing relation 
    $\mathtt{has\_type} : \mathtt{Expr} \times \mathtt{Ty} \rightarrow \mathtt{Prop}$.
  \item\label{enum:lab:algo} Typechecking algorithm: 
    $\mathtt{type\_check} : \mathtt{Expr} \rightarrow \mathtt{option\ Ty}$.
  \item\label{enum:lab:proof} Proof of correctness\footnote{Completeness condition ($\Leftarrow$) does not always hold.}:\\ 
    $\forall e, \tau.\mathtt{has\_type}(e, \tau) \Leftrightarrow \mathtt{type\_check}(e) = \mathtt{Some}\ \tau. $
\end{enumerate}

Whereas such tasks as typechecking depend a lot on
a programming language,
it seems that there are certain
parts of interpreter 
that can be implemented in an \femph{abstract} manner.
Thus, for example, most of the mainstream programming languages 
have some notion of module.
Or even more broadly, a notion of list of declarations/definitions. 
It could be a list of class declarations in a package,
method declarations in an interface, 
type and function definitions in a module, etc.
The well-definedness condition for a list of declarations
can be \femph{abstractly} formulated as follows:
\begin{enumerate}
  \item All names in the list are different.
  \item\label{enum:lab:decls} Every declaration in the list is well-defined.
\end{enumerate}
To get a concrete well-definedness property for a particular
module-like construct, 
it suffices to substitute abstract parts, such as ``name'' 
or ``well-defined declaration'', with concrete types and predicates.

We suggest that a substantial part of 
propositions (layer~\ref{enum:lab:prop}),
algorithms (layer~\ref{enum:lab:algo}), and proofs (layer~\ref{enum:lab:proof})
related to certified checking of module-like language constructs
can be implemented in abstract way, as a generic Coq library.
Partial implementation of the library is available in GitHub~\cite{bib:ConceptParams-Coq}.
We provide two sorts of generic code.
The first one is more 
low-level, related to the efficient
representation of finite maps (see more in \sectref{sec:efficiency}). 
The second one is more high-level, connected with the semantics 
of modules (\sectref{sec:modules}).

\section{Efficiency Matters}\label{sec:efficiency}

One subtlety in building certified interpreters is efficiency.
Note that formal models are aimed to \femph{reason} about 
programming languages but are not supposed to run.
Therefore, in particular, there is no need to use efficient
data structures for representation of programs, contexts, or types.
Interpreters, by contrast, are to be \femph{executed}. 
Therefore, they better use efficient data structures
and algorithms to represent, analyse, and run programs.
However, efficient code could be harder to reason about.
Furthermore, as we want to certify an interpreter against a model,
it means that the model and the interpreter could share some code.
This, in turn, leads to more sophisticated reasoning about the model itself.
Thus, there is a conflict between efficiency and ease of reasoning.

In the context of module-like constructs,
an example of such a conflict is the representation of finite maps.
In the first place a program
is given in the form of an abstract syntax tree (AST).
If such an AST contains a well-defined list of declarations,
an interpreter can convert the list to
a finite map from names (identifiers) to some data.
Or, alternatively, it can further use the AST as is.
The latter way is less efficient but more straightforward, 
as no extra proofs are needed to show
that the result finite map is ``equivalent'' to the source AST.
That is why this approach is normally used in mechanised formal 
models~\cite{bib:Delaware:2011:PLT:2076021.2048113,bib:Rompf:2016:TSD:3022671.2984008,bib:Mackay:2012:EFJ:2318202.2318206}
to describe records, classes, and namespaces. 
By contrast, the CompCert compiler uses efficient 
tree-based finite maps for representation of programs
(a program is defined as a list of function and variable declarations).

Specifically, if there is a function \fcode{map\_from\_list}
which converts a list of declarations into a finite map, 
one has to prove a bunch of properties about it.
For example, assuming that an AST of declarations list 
is represented by list of pairs \texttt{(<name>, <data>)}, 
it must be proven that 
$ \forall n, d . [n \mapsto d] \in (\mathtt{map\_from\_list}\ \mathit{decls}) 
\Rightarrow (n, d) \in \mathit{decls}. $
Such kind of properties are proven in our library for
a transformation of a list of pairs 
into a generic interface of finite maps 
\fcode{FMap} (from the Coq standard library\footnote{CompCert does
some similar things for its own interface of finite maps.}).
There are some other proven-to-be-correct functions,
e.g. generic \fcode{ids\_are\_unique}, which checks
repetitions in a list using an auxiliary set.

\section{Modules}\label{sec:modules}

As we mentioned in~\sectref{sec:intro}, many programming languages support some kind
of module-like constructs that introduce namespaces.
But what is more important, ``modules'' provide an instrument of abstraction~---
they allow for separation of interface from implementation.
Examples could be interfaces and classes in Java, protocols and classes in Swift,
signatures and modules in ML, type classes and instances in Haskell.
The main difference between a module-interface and a module-implementation
is that the former one must be well-defined,
and the latter one must be well-defined \femph{with respect to} the former one.
Although, in presence of structural subtyping, well-definedness of a module-interface 
can also depend on some other interfaces.

Our ultimate goal is a generic Coq library, which provides building blocks
for certified checking of modules of different flavours.
For instance, compare Java~7 and Java~8 interfaces.
The latter support default method implementations, while the former do not.
It means that Java~7 class, which extends an interface, 
is well-defined only if all interface member are defined. 
But Java~8 class is well-defined under the relaxed condition,
if all not-implemented interface members are defined.
Another difference in presence of default implementations is
a way method declarations are checked (part \ref{enum:lab:decls} of our
well-definedness property). 
In Java 8 interfaces, method bodies can refer to other 
methods of the same interface, whereas in Java 7 
there is no need to take into account a local context.
A bit of a different approach is needed for ML signatures/modules,
where declarations can only refer to previously-defined ones.
One more variation of well-definedness is required 
for mutually recursive definitions.

Following the structure of certified interpreter, our library
consists of the triples: propositional definitions of well-definedness,
computable functions for checking well-definedness, 
proofs of correctness. 
Every part of a triple is a functor parameterized over
type of identifiers, decidable equality of identifiers, 
type of data, type of context, and some other things.
As an example, consider the simplest possible semantics of module-interfaces,
where all declarations can be checked independently of each other.
Assuming that an interface is given as a list of pairs
\fcode{(id, ty)}, a ``propositional'' functor might look as follows:
\begin{cpSTLCNoFrame}
Module SimpleIntrfs_Defs (Import ...).
  Definition types_ok (c : ctx) (tps : list ty) : Prop :=
    List.Forall (fun tp => is_ok c tp) tps.
  Definition module_ok (c : ctx) (ds : list (id * ty)) : Prop :=
    let (nms, tps) := split ds in
    (** all names are distinct, all types are well-defined *)
    List.NoDup nms /\ types_ok c tps.     
\end{cpSTLCNoFrame}
\vspace{-0.5em}
In addition to other parameters, \fcode{SimpleIntrfs\_Defs} depends on
the proposition $\mathtt{is\_ok} : \mathtt{ctx} \rightarrow \mathtt{ty} \rightarrow \mathtt{Prop}$,
which defines what it means for a type to be well-defined 
in the given global context. 
A functor with computable functions is defined in a similar way and
implements functions \fcode{types\_ok\_b} and \fcode{module\_ok\_b},
which return \fcode{bool}. 
Finally, there is a proofs functor, which
proves that the computable functions are correct 
with respect to the propositions. 


We justify this generic implementation by applying it
to an extension of simply typed lambda calculus with simple
modules~--- concepts and models, 
which is proven to be type sound.
``Concept'' represents module-interface, it consists of
name-type pairs. 
``Model'' represents module-imple\-men\-ta\-tion for a particular
concept: it consists of name-term pairs, 
with terms referring to the previously defined ones
and having types declared in the concept.
Terms are terms of STLC extended with 
module-related constructs:\\
(1) Concept abstraction $\lambda c\mathbin{\#}\mathtt{C}.\,e$,
    which allows $e$ to refer to the members of concept $\mathtt{C}$ 
    via concept variable $c$. \\
(2) Member invocation $c{::}f$. \\
(3) Model application $e \mathbin{\#}\mathtt{M}$, which is valid
    only if $e$ is a concept abstraction $\lambda c\mathbin{\#}\mathtt{C}.\,e'$,
    with $\mathtt{M}$ being a model of $\mathtt{C}$.\\
Typing of terms is a five-place relation, which takes into account
contexts of concepts and models:
$\texttt{CT} * \texttt{MT} ; \Gamma \vdash t : \tau.$
Contexts $\texttt{CT}$ and $\texttt{MT}$ are required to be well-defined.
We use our generic library four times to typecheck a program
in this language. Namely, we use it to check 
a single concept/model definition, 
a section of concept definitions, and a section of model definitions.
More complicated strategies of dealing with modules is a subject 
for future work.


\bibliographystyle{ACM-Reference-Format}
\bibliography{belyakova-biblio} 

\end{document}